\documentclass[preprint,preprintnumbers,amsmath,amssymb]{revtex4}

\usepackage{graphicx,amsbsy,dcolumn}
\usepackage{amsmath,amsfonts,euscript,epsfig,color}
\usepackage{mathrsfs} 
\DeclareMathAlphabet{\mathbsf}{OT1}{pss}{sbc}{sl}
\newcommand{\bm}[1]{\mbox{\boldmath $#1$}}     
\newcommand{\mfrac}[2]{{\textstyle\frac{#1}{#2}}}
\newcommand{\ncien}[2]{\mbox{$#1\,\text{E$#2$}$}}
\newcommand{\uniC}{\mbox{i}}
\def\re{\mathbb{R}}
\newcommand{\dpar}[2]{\frac{\partial #1}{\partial #2}}
\newcommand{\cte}{\mbox{const\,}}
\newcommand{\grad}{\mbox{grad\,}}
\newcommand{\diag}{\mbox{diag\,}}
\def\sen{\mathop{\rm sen}\nolimits}
\begin{document}

\title[Curvature of a discretized light wavefront]{Numerical estimation of the curvature of a light wavefront in a weak gravitational field}

\author{A. San Miguel, F. Vicente and J.-F. Pascual-S\'anchez}

\address{Dept. de Matem\'atica Aplicada,
Facultad de Ciencias.\\
Universidad de Valladolid, 47005 Valladolid, Spain\\}

\begin{abstract}
The geometry of a light wavefront evolving in the 3--space associated with a post-Newtonian relativistic spacetime from a flat wavefront is studied numerically by means of the ray tracing method. For a discretization of the bidimensional wavefront the surface fitting technique is used to determine the curvature of this surface at each vertex of the mesh. The relationship between the curvature of a wavefront and the change of the arrival time at different points on the Earth is also numerically discussed.
\end{abstract}

\pacs{04.30.Nk, 02.60.Cb, 04.25.Nx}

\maketitle

\section{Introduction} \label{intro}
The description of the propagation of light in a gravitational field is even today a central problem in the general theory of relativity. The deflection of light rays and time delays of electromagnetic signals due to the presence of a gravitational field are phenomena detectable with current experimental techniques which allow design new tests for general relativity. In this line, Samuel~\cite{Sam} recently proposed a method for the direct measurement of the curvature of a light wavefront initially flat, curved when light crosses regions where the gravitational field is non vanishing. He found a relationship between the differences of arrival time recorded at four points on the Earth, measured by employing techniques of very long base interferometry and the volume of a parallelepided determined by four points in the curved wavefront surface. This surface is described by means of a polynomial approximation of the eikonal in a Schwarzschild gravitational field. For more complex gravitational models, such as those considered by Klioner and Peip~\cite{KP}, de Felice {\it et al.}~\cite{dF} or Kopeikin and Sch\"afer~\cite{KS} in studies of light propagation in the solar system, the use of numerical methods for the determination of the geometry of the wavefront surface would also be required. An analytical approach to the relativistic modeling of light propagation has also been developed recently by Le Poncin-Lafitte {\it et al.}~\cite{Lep} and Teyssandier and Le Poncin-Lafitte~\cite{Tey}, where they present methods based on Synge's world function and the perturbative series of powers of the Newtonian gravitational constant, to determine the post-Minkowskian expansions of the time transfer functions.

Nowadays there are numerous techniques in computational differential geometry which allow to analyze geometric properties of surfaces embedded in the ordinary Euclidean space. Techniques of this type are widely applied in different areas such as Computational Geometry, Computer Vision or Seismology. In one of these methods, developed in works by Garimella and Swartz~\cite{GS} and  Cazals and Pouget~\cite{CP}, the estimation of differential quantities is established using a fitting of the local representation of the surface by means of a height function given by a Taylor polynomial. A survey of methods for the extraction of quadric surfaces from triangular meshes is found in Petitjean~ \cite{Pet}.

In this work we consider a discretization of the wavefront surface, replacing this surface by a polyhedral whose faces are equilateral triangles. At initial time, the surface is assumed to be flat and far enough from a gravitational source (say, the Sun) and moving towards this source. We study the deformation of the instantaneous polyhedral representing the wavefront when crossing a region in the relativistic 3--space near the Sun due to the bending of light rays by the gravitational field. In this study, we apply the ray tracing method with initial values on the vertices of the triangular mesh to obtain the corresponding discrete surface at each instant of time. Then we apply the techniques given in \cite{GS} and \cite{CP} to describe the wavefront as a surface embedded in the Riemannian 3--space of the post-Newtonian formalism of general relativity. For each vertex in the instantaneous mesh we obtain a quadric which represents locally the surface by applying the least-squares method to the immediate neighboring vertex around the considered point which is represented in normal coordinates adapted to the light rays.

The structure of the paper is as follows: In Section 2 we briefly introduce the basic model for the wavefront propagation in the post-Newtonian formalism. In Section 3, we establish a discretized model of the wavefront surface by means of a regular triangulation and we describe the method employed in this work for the study of the curvature of this surface. In Section 4, a numerical estimation of the curvature of the surface is derived using the ray tracing method. For the numerical integration of the light ray equation, we use the {\tt Taylor} algorithm implemented by Jorba and Zou \cite{JZ} which is based on the Taylor series method for the integration of ordinary differential equations and which allow the use of high order numerical integrators and arbitrary arithmetic accuracy, as is required to describe the influence of weak gravitational perturbations on the bending of light rays. Finally, in Section 5, we apply the method discussed above to study the effect of the wavefront curvature on the variation of the arrival time of the light at points on the Earth surface, following the model proposed in Samuel's test \cite{Sam}. The paper concludes by giving another approach to the estimation of the curvature of the wavefront, derived from an approximation of the Wald curvature \cite{Blu} associated with a quadruple of points in the wavefront.

\section{Light propagation in a gravitational field}
Let us consider a spacetime $(\mathscr{M},g)$ corresponding to a weak gravitational field and choose a coordinate system  $\{(\bm{z},ct)\}$ such that the coordinate representation of the metric tensor is
\begin{equation}\label{tensmetr}
    g_{\alpha\beta}=\eta_{\alpha\beta}+h_{\alpha\beta}, \qquad\text{with}\quad \eta_{\alpha\beta}=\diag(1,1,1,-1).
\end{equation}
where the coordinate components of the metric deviation $h_{\alpha\beta}$ are given by:
\begin{equation}\label{PN4}
h_{ab} =  2c^{-2}\kappa \|\bm{z}\|^{-1}\delta_{ab},\qquad
h_{a4} = -4c^{-3}\kappa \|\bm{z}\|^{-1}\dot{Z}_a, \qquad
h_{44} =  2c^{-2}\kappa \|\bm{z}\|^{-1}
\end{equation}
(Greek indices run from 1 to 4 and Latin indices from 1 to 3) where $\kappa:=GM$ represents the gravitational constant of a monopolar distribution of matter (say the Sun) located at $Z^a(t)$ and $c$ represents the light speed.

In the post-Newtonian framework one may consider a simultaneity space $\Sigma_t$ at each coordinate time $t$. From the fundamental equation of the geometrical optics  for the phase $\psi(z,t)$ of an electromagnetic wave \cite{LL}:
\begin{equation}\label{iconal}
g^{\alpha\beta}\dpar{\psi}{z^\alpha}\dpar{\psi}{z^\beta}=0,
\end{equation}
and Cauchy data $\psi(z,0)=\text{const}$, given on a spacelike surface $\mathscr{D}_0:=\{(z,0)\,|\; \phi(z)=0\}\subset\Sigma_0$ one obtains the characteristic hypersurface (light cone) $\Omega:=\{(z,t)\,|\; \psi(z,t)=\cte\}\subset \mathscr{M}$ of the light propagation. The intersection of the characteristic hypersurface  and the corresponding simultaneity space is the spacelike wavefront at a time $t$ which will be denoted by $\mathscr{S}_t:=\Omega\cap\Sigma_t$.

An alternative formulation of the problem of light propagation in the spacetime $(\mathscr{M},g)$ is based on the determination of the bicharacteristics generated by the isotropic vectors $k:=\grad \psi$. From both the equation for the null geodesics, $z(t)=\big(\bm{z}(t),t\big)$ expressed in terms of the coordinate time and the isotropy condition $g(\dot{z},\dot{z})=0$, one obtains in the post-Newtonian approach to general relativity that, neglecting terms of order $O(c^{-2})$, the null geodesics of $(\mathscr{M},g)$ must satisfy the equations
\begin{eqnarray}
        \ddot{z}^a &= \varphi_a(\bm{z},\dot{\bm{z}},t) \label{PN1a}\\
         0 &= g_{\alpha\beta}\dot{z}^\alpha\dot{z}^\beta.\label{PN1b}
\end{eqnarray}
In (\ref{PN1a}) the components of the acceleration $\varphi_a(\bm{z},\dot{\bm{z}},t)$ are given by (see \cite{Bru})
\begin{eqnarray}\label{PN2}
    \varphi_a(\bm{z},\dot{\bm{z}},t) = &
    \mfrac{1}{2} c^2h_{44,a} -[\mfrac{1}{2}h_{44,t}\delta^a_k+h_{ak,t}+c(h_{4a,k}-h_{4k,a})]\dot{z}^k\nonumber\\
     & -(h_{44,k}\delta^a_l+h_{ak,l}-\mfrac{1}{2}h_{kl,a})\dot{z}^k\dot{z}^l \nonumber\\
     & -(c^{-1}h_{4k,j}-\mfrac{1}{2}c^{-2}h_{jk,t})\dot{z}^j\dot{z}^k\dot{z}^a,
\end{eqnarray}
where the first and third terms in the right hand side of (\ref{PN2}) are of order $O(1)$, while the remaining terms are $O(c^{-1})$.

For initial values  $(\bm{z}_0,\dot{\bm{z}}_0)$, with $\bm{z}_0\in\mathscr{S}_0$ and $(\dot{\bm{z}}_0/c,1)$, satisfying the condition (\ref{PN1b}),  the integration of the initial value problem  corresponding to (\ref{PN1a}) on an interval $[0,T]$ allows to determine the spacelike wavefront $\mathscr{S}_T$. This surface is embedded in the Riemannian 3--space $(\Sigma_T,\tilde{\gamma})$  where the components of the metric tensor are given by
\begin{equation}\label{trimetr}
\tilde{\gamma}_{ab}:=g_{ab}-\frac{g_{a4}g_{b4}}{g_{44}},
\end{equation}
and at the time $T$ the tangent vectors $\dot{\bm{z}}(T)$ to the integral curves $\bm{z}(t)$ are $\tilde{\gamma}$--orthogonal to $\mathscr{S}_T$.

\section{Numerical description of a spacelike bidimensional wavefront}\label{sec3}
Hereafter, we consider the simplest gravitational model generated by a static point mass. Let $\mathscr{E}$ denote the quotient space of $ \mathscr{M} $ by the global timelike vector field  $ \partial_t$ associated with the global coordinate system used in the post-Newtonian formalism.  We will consider a region of the wavefront in $\mathscr{E}$ described by a coordinate chart  $\{z\}$. Further, we assume that the Riemannian manifold $(\mathscr{E},\bm{\tilde{\gamma}})$ is almost flat and that the metric corresponding to $\bm{\tilde{\gamma}}$  is quasi-Cartesian in the chosen coordinates.

\subsection{Discretization of the initial wavefront}\label{subsecDiscFO}
Given an asymptotically Cartesian coordinate system, we consider a set $\mathscr{S}_0$ formed by points with coordinates $(z_1,z_2,-\zeta)$ where $\zeta>0$ is a number large enough so that $\mathscr{S}_0$ may be considered as a flat surface. The direction determined by the point $O^*:(0,0,-\zeta)$ in $\mathscr{S}_0$ and the center of the Sun $O:(0,0,0)$ is  perpendicular to $\mathscr{S}_0$. We will study the geometry of a region $\mathscr{C}\subset \mathscr{S}_0$ determined by points $P$ whose Euclidean distances $d(P,O^*)$ to the point $O^*$ satisfy $R_{\odot}\leqslant d(P,O^*)\leqslant 2R_{\odot}$, where $R_\odot$ is the radius of the Sun.

\begin{figure}
\begin{center}
  \includegraphics[width=10cm,angle=0]{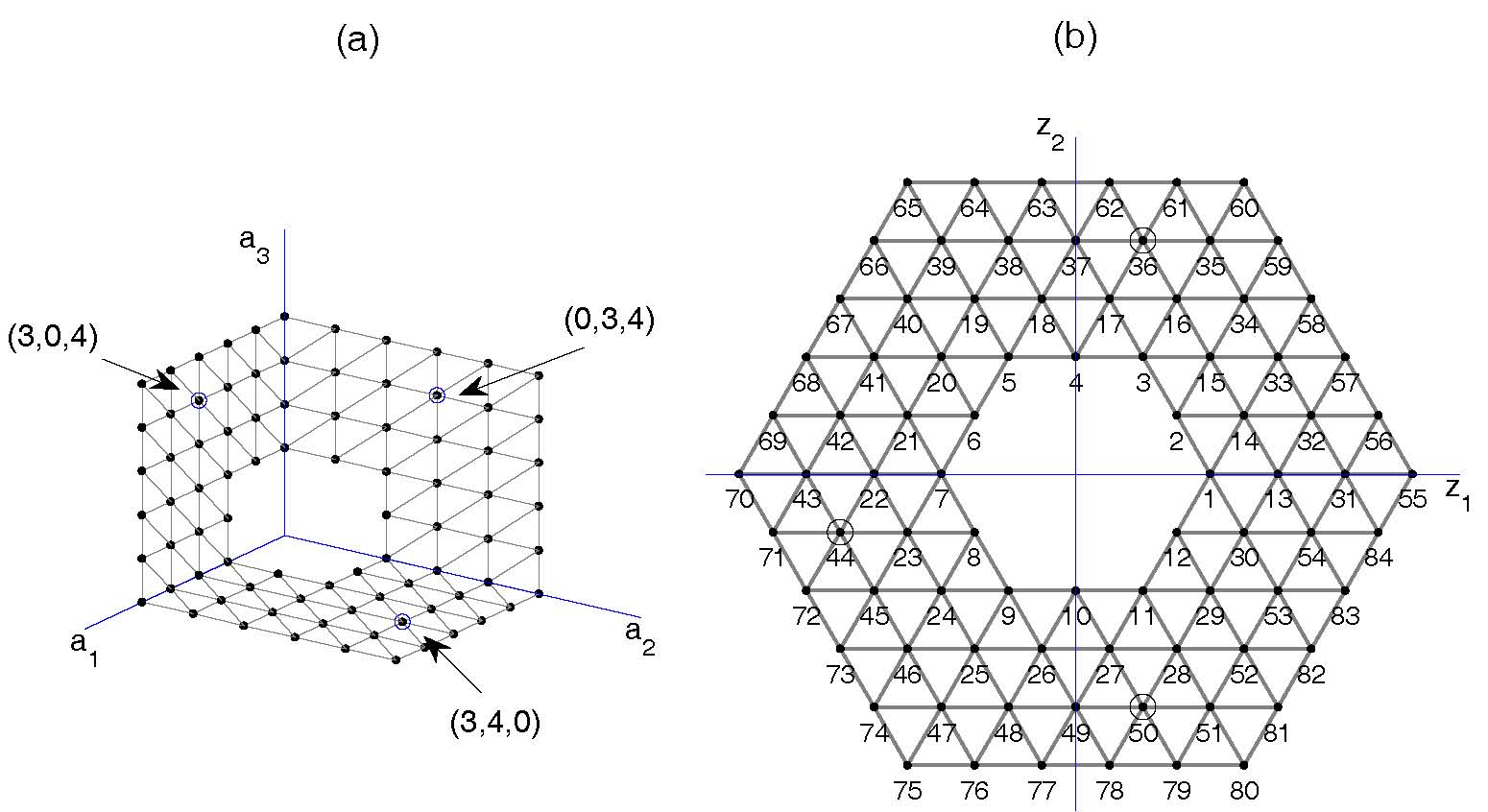}
  \caption{Triangulation and enumeration of the initial wavefront. (a) Triplets $(a_1,a_2,a_3)\in \mathscr{A}^{*3}$ used to label the vertex of a hexagonal mesh $\bar{\mathscr{V}}$. (b) Regular triangulation $\mathscr{V}^*$ of a hexagonal annular region in the wavefront. Points $(3,4,0), (0,3,4)$ and $(3,4,0)$ in $\bar{\mathscr{V}}$ correspond with points $36, 44$ and $50$ in $\mathscr{V}^*$ respectively.} \label{figure1}
\end{center}
\end{figure}

For the discretization of the problem we consider, in the first place, the set of points $(a_1,a_2,a_3)\in\mathscr{A}^{*3}$, where $\mathscr{A}=\{0, N_1, N_1+1, \dots, N_2\}$ and $N_1<N_2$ are two natural numbers (see Figure ~\ref{figure1}(a)). In $\mathscr{A}^{*3}$ the point $(0,0,0)$ is excluded.
Next, we construct in the complex plane a regular triangular mesh whose vertices are located between two hexagons as shown in figure~\ref{figure1}(b), and the edges have length $\ell$. The inner hexagon has sides of length $\ell N_1$ and the outer hexagon sides of length $\ell N_2$. In this triangulation each vertex is represented by a complex number of the  set (see \cite{BH})
\begin{equation}\label{vertices1}
    \bar{\mathscr{V}}:=\{z=a_1+a_2\omega+a_3\omega^2\;|\;\; a_1,a_2,a_3\in\mathscr{A},\; \omega:=\exp(2\pi \uniC/3)\},
\end{equation}
where $\uniC :=\sqrt{-1}$. The vertices $(a_1,a_2,a_3)$ with some of their components equal to $N_1$ (resp. $N_2$) are located on the inner (resp. outer) boundary  of the mesh $\bar{\mathscr{V}}$. We establish an enumeration of the vertices as shown in Figure~\ref{figure1}(b), in such a way that the inner vertices $z_j$ have subscripts   $j=1,\dots, J$. Finally, we apply the change of scale:
\begin{equation}\label{vertices2}
    \bar{\mathscr{V}}\rightarrow\mathscr{V}^*, \qquad z\mapsto \frac{z r}{N_1},
\end{equation}
so that the inner boundary is a hexagon of radius $r$. In consequence the length of each edge in this triangulation is equal to $r/N_1$.

The complex plane and the plane $\mathscr{S}_0$ may be identified by means of the mapping $\iota:z\mapsto (\Re(z),\Im(z),-\zeta)$. Thus one obtains a discretization of the initial region of $\mathscr{S}_0$ which we are considering here, triangulated with vertices given by $\iota(z_j)$ whose corresponding mesh on $\mathscr{S}_0$ will also be denoted by $\mathscr{V}^*$.

\subsection{Normal coordinates around a point}
Now we assume that at each vertex in $\mathscr{V}^*$ a photon with velocity $\dot{\bm{z}}_0:=(0,0,c)$ is located. The null geodesics equation (\ref{PN1a}) may be written as a first order differential system $\dot{\bm{u}}=\bm{F}(\bm{u},t)$ in phase space $\bm{u}:=(\bm{z},\dot{\bm{z}})$ which determines a flow in $\mathscr{E}$:
    \begin{equation}\label{flow}
    \bm{z}(t) =\varPhi_t(\bm{z}_0,\dot{\bm{z}}_0),
    \end{equation}
in terms of the initial values $\bm{z}_0:=\bm{z}(0)\in\mathscr{V}^*$, $\dot{\bm{z}}_0:=\dot{\bm{z}}(0)$. Then, for each time $t$ there is a surface $\mathscr{S}_t$ image of $\mathscr{S}_0$ under the flow (\ref{flow}). At a point $\bm{z}\in\mathscr{S}_t$ the normal vector $\bm{\tau}(\bm{z})$ coincides with the tangent vector to the curve $\bm{z}(t)$ at that point. Furthermore, before reaching the focal points of the beam of light, the triangulation $\mathscr{V}^*$ induces a triangulation $\mathscr{V}$ on the wavefront $\mathscr{S}_t$  whose vertices we enumerate using the same labels used for the corresponding vertices in $\mathscr{S}_0$.

In order to simplify the description of the geometry of the surface $\mathscr{S}_t$ on a neighborhood of a point $P\in \mathscr{S}_t$ we use a $\bm{\tilde{\gamma}}$--orthonormal reference frame $\{\bm{e}_i\}_{i=1}^3$ centered on that point, where one of its vectors, say $\bm{e}_3 $, is parallel to the vector    $\bm{n}(P):=\bm{\tau}(P)/\sqrt{\tilde{\gamma}(\bm{\tau},\bm{\tau})}$ tangent to the ray passing through that point. Let  $\{y^j\}_{j=1}^3$ be a normal coordinate system with pole at the point $P$ and associated normal reference frame $\{\bm{e}_i\}_{i=1}^3$ .

By using the classic formulae of Riemannian geometry (see \cite{Eis} \S 18, and \cite{Br}), the coordinate transformation $y^i\mapsto z^i$,  from normal to post-Newtonian coordinates, is determined by
\begin{equation}\label{1}
    y^i= (\Lambda^{-1})^i_a\big(z^a-z^a_0 + \mfrac{1}{2}\big(\tilde{\Gamma}^a_{bc}\big)_0(z^b-z^b_0)(z^c-z^c_0)\big),
\end{equation}
where $\Lambda$ is a non-singular constant matrix and $\big(\tilde{\Gamma}^a_{bc}\big)_0:=\tilde{\Gamma}^a_{bc}(P)$ are the Christoffel symbols at the point $P$. By neglecting terms of order higher than $\tilde{\Gamma}^a_{bc}$, the inverse transformation may be approximated by
\begin{equation}\label{1_}
    z^a= a_0^a+\Lambda^a_i\big(y^i - \mfrac{1}{2}\big(\tilde{\Gamma}^i_{jk}\big)_0y^jy^k\big),
\end{equation}
whose corresponding Jacobian determinant is
\begin{equation}
    \dpar{z^a}{y^i}=\Lambda^a_k\big(\delta_i^k-\big(\tilde{\Gamma}^k_{il}\big)_0y^j\big).
\end{equation}
In  normal coordinates a metric tensor $\gamma$ on the space $\mathscr{E}$  is determined from $\tilde{\gamma}$ by
\begin{equation}\label{trimetricNor}
    \gamma_{ij} =\dpar{z^a}{y^i}\dpar{z^b}{y^j}\tilde{\gamma}_{ab}.
\end{equation}
and, as it is well known, at point $P$ the tensor $\gamma_{ij}$ is reduced to $\delta_{ij}$  and the associated Christoffel symbols at this point are $\Gamma^i_{jk}=0$.

\subsection{Local approximation of the wavefront}\label{ALFO}
From the triangulation $\mathscr{V}^*$ of the initial wavefront the ray tracing method furnishes a discrete surface determined by the mesh $\mathscr{V}$. To compute differential magnitudes of the wavefront surface corresponding to this mesh one needs to define a discrete neighborhood of each vertex in $\mathscr{V}$. For this, we consider firstly
at each inner vertex $z_j^*\in \mathscr{V}^*, j=1,\dots,J$ a neighborhood (named hereafter {\sl 1--ring} in the terminology of computational geometry, {\it e.g.} \cite{MDSB}) formed by the six vertices $z^*_{j_k}$ closest to $z^*_j$:
\begin{equation}\label{1anillo}
    [z^*_j;z^*_{j_k}]_{j=1,\dots,J,k=0,\dots,5},\qquad \text{where}\quad z^*_{j_k}:=z_j+\exp(k\pi\uniC/3),
\end{equation}
Then, for each 1--ring in (\ref{1anillo}) one may determine on the mesh $\mathscr{V}$ a corresponding 1--ring formed by the image of the points $z^*_j,z^*_{j_k}\in\mathscr{S}_0$ under the flow (\ref{flow})
\begin{equation}\label{EntornoD}
    [z_j;z_{j_k}] := [\varPhi_t(z^*_j); \varPhi_t(z^*_{j_k})].
\end{equation}

In a neighborhood of a point $z_j$ the wavefront can be approximated by a height function on the orthogonal plane  to $\bm{e}_3$ determined by means of a least-squares fitting of the data (\ref{EntornoD}) expressed in normal coordinates $\{y^j\}_{i=1}^3$. As a model for this surface we chose a quadric passing through the coordinate origin whose gradient at this point is parallel to $\bm{e}_3$,
\begin{equation}\label{2}
    y^3=f(y^1,y^2):=\mfrac{1}{2}a_1(y^1)^2 + a_2y^1 y^2 + \mfrac{1}{2}a_3(y^2)^2,
\end{equation}
where $ a_1, a_2$ and $a_3$ are the indeterminate coefficients to be obtained by the least squares method.

The quadric (\ref{2}) provides a surface  $\tilde{\mathscr{S}}_j$ that approximates the surface $\mathscr{S}_t$ in a neighborhood of the point $z_j$ and is defined in parametric form $y^i=y^i(x^A)$, with $A=1,2$, as
\begin{equation}\label{cuadrica}
    y^1=x^1, \quad y^2=x^2,\quad y^3=f(x^1,x^2).
\end{equation}
In coordinates $(x^1,x^2)$, the metric $\bm{\gamma}$ on $\mathscr{E}$ induces a metric on $\tilde{\mathscr{S}}_j$ whose associated metric tensor $\bm{g}$ has the form
\begin{equation}\label{metrind}
    g_{AB} := \gamma_{ij}\dpar{y^i}{x^A}\dpar{y^j}{x^B}.
\end{equation}
Moreover, on the tangent plane to $\mathscr{S}_t$ at $z_j$ one defines a tensor $B$ associated with the normal  $\bm{n}$ at that point as:
\begin{equation}\label{B2}
    B:T_{z_j}\mathscr{S}\times T_{z_j}\mathscr{S}\rightarrow (T_{z_j}\mathscr{S})^\perp,\qquad B(\partial_A,\partial_B)=\frac{\partial^2y^3}{\partial x^A\partial x^B}\bm{n}
\end{equation}
Therefore, if $\{\bm{v}_1,\bm{v}_2\}$ represents an orthonormal basis for the vector space $T_{z_j}\mathscr{S}$ consisting of eigenvectors of $B$ with  associated eigenvalues $\lambda_1$ and $\lambda_2$, and $I\!I$ denotes the second fundamental form, then the difference of sectional curvatures $K$ and $\bar{K}$ associated with the plane generated by $\{\bm{v}_1,\bm{v}_2\}$, in $\mathscr{S}$ and $\mathscr{E}$ respectively, is given by a generalized Gauss formula (see \cite{dC}, p.131)
\begin{equation}\label{FGt}
    K_{\text{rel}}:=K(\bm{v}_1,\bm{v}_2)-\bar{K}(\bm{v}_1,\bm{v}_2)=\lambda_1\lambda_2
\end{equation}
which is named (see \cite{Eis}) relative sectional curvature $K_{\text{rel}}$, whereas the mean curvature is determined by half the trace of $I\!I$:
\begin{equation}\label{FGm}
    H=\mfrac{1}{2}(\lambda_1+\lambda_2)
\end{equation}

\section{Numerical estimation of the curvature of a wavefront}
\subsection{Integration of the equations of light rays}\label{subsec41}
Here we deal with the problem of the numerical integration of the initial value problem corresponding to (\ref{PN1a})-(\ref{PN1b}). The equation (\ref{PN1a}) for the light propagation in the gravitational field generated by a static material point, may be rewritten in terms of new variables defined as:
\begin{equation}\label{varu}
    \bm{u}:=(u_1,\dots, u_6),\quad \text{with $u_i:=z_i$ and  $u_{i+3}:=\dot{z}_i,\quad  (i=1,2,3)$,}
\end{equation}
obtaining the following first order differential system:
\begin{equation}\label{edo2}
\begin{split}
\dot{u}_1 &= u_4,\\
\dot{u}_2 &= u_5,\\
\dot{u}_3 &= u_6,\\
\dot{u}_4 &= -\frac{\kappa}{c^2}\frac{c^2 u_1 -3u_1u_4^2-4u_2u_4u_5-4u_3u_4u_6+u_1u_5^2+u_1u_6^2}{(u_1^2+u_2^2+u_3^2)^{3/2}},\\
\dot{u}_5 &= -\frac{\kappa}{c^2}\frac{c^2 z_2 -4u_1u_4u_5-3u_2u_5^2-4u_3u_5u_6+u_2u_4^2+u_2u_6^2}{(u_1^2+u_2^2+u_3^2)^{3/2}},\\
\dot{u}_6 &= -\frac{\kappa}{c^2}\frac{c^2 z_3 -4u_1u_4u_6-4u_2u_5u_6-3u_3u_6^2+u_3u_4^2+u_3u_5^2}{(u_1^2+u_2^2+u_3^2)^{3/2}}.
\end{split}
\end{equation}
The solution $\bm{u}(t)$ must satisfy the constraint (\ref{PN1b}) which, in the notation (\ref{varu}), may be expressed as
\begin{equation}\label{conlig}
F(\bm{u}):=u_4^2+u_5^2+u_6^2-c^2+\frac{2\kappa}{c^2\sqrt{u_1^2+u_2^2+u_3^2}}(u_4^2+u_5^2+u_6^2+c^2)=0.
\end{equation}

To obtain a numerical  solution of the initial value problem given by (\ref{edo2}) and the initial data $\bm{u}(0)=\big(\bm{z}(0),\dot{\bm{z}}(0)\big)$ we use the {\tt Taylor} integrator developed by Jorba and Zhou \cite{JZ}, based on the classic Taylor series method for ordinary differential equations. In this method a Taylor expansion of the vector field $\dot{\bm{u}}$ defined in (\ref{edo2}) is made using techniques of automatic differentiation to obtain the corresponding Taylor coefficients. The {\tt Taylor} integrator allows the control of both the order and the step size employed in the method. Furthermore, the {\tt Taylor} integrator is implemented so that one may use extended precision arithmetic for the highly accurate computation required in this problem.

\begin{figure}
\begin{center}
  \includegraphics[width=9cm,angle=0]{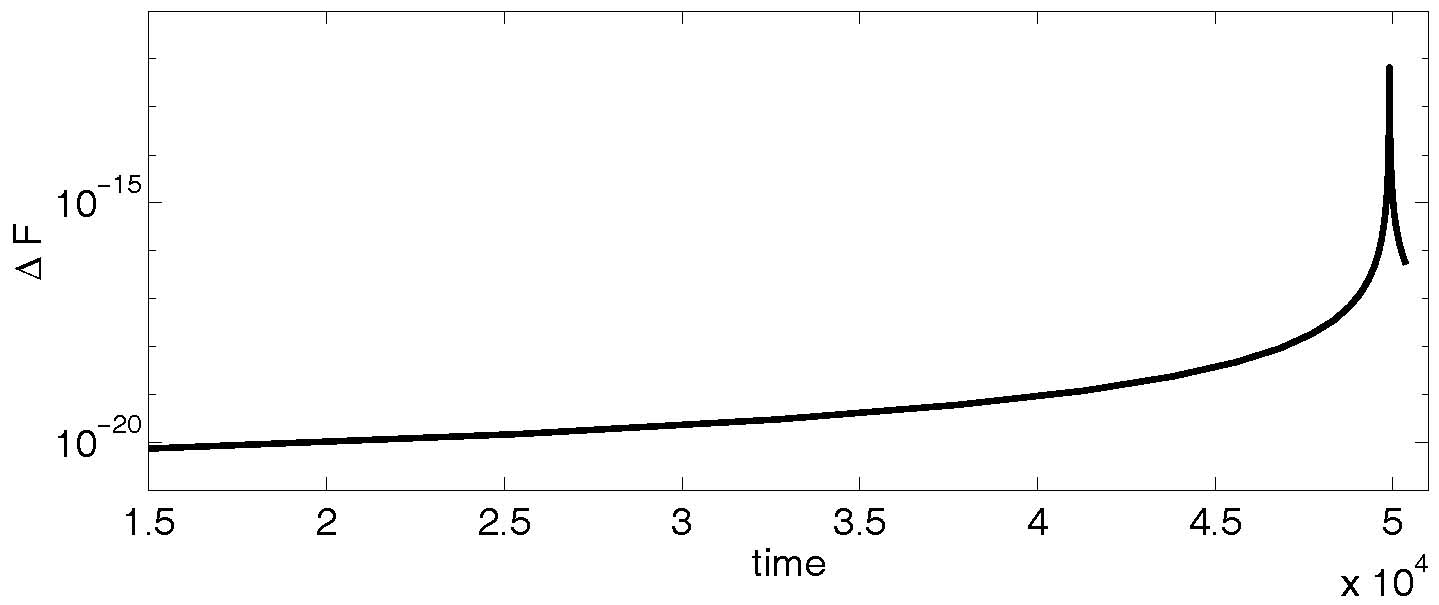}
  \caption{Semilogarithmic representation of the difference $F(\tilde{u}_n)-F(u_0)$ along the numerical solution of the light ray equation obtained by means of the {\tt Taylor} algorithm.  The maximum value of $\Delta F$ is reached at the time when the photon is nearest to the Sun.} \label{figure2}
\end{center}
\end{figure}

From now on we use normalized units taking the radius and the mass of the Sun as units of length and mass, respectively. Then, the initial values corresponding to a photon initially located at 100 astronomical units from the Sun are given by
\begin{equation}\label{coninic}
    \bm{u}_0:=\bm{u}(0)=(1.0, 0.0, -21494.6550, 0.0, 0.0, 0.4307).
\end{equation}
For the integration over a period of time of $50400$ seconds, using a precision of $120$ binary digits and a tolerance $\text{\tt Tol}=\ncien{1.}{-20}$, the {\tt Taylor} algorithm gives the solution in terms of Taylor polynomials of degree twenty-four. As a test of the accuracy of the method, in Figure~\ref{figure2} it is shown the behavior of the function $F(\bm{u})$ appearing in the isotropy constraint (\ref{PN1b}). In this figure we see that at the arrival point, the isotropy constraint on the tangent vector to the light ray is not satisfied with the same degree of accuracy required at the initial position, reaching this deviation its maximum value at the instant when the photon is nearest to the Sun.

In order to obtain from the numerical solution $\tilde{\bm{u}}_n$ given by {\tt Taylor} another value $\bm{u}_n$  satisfying the condition (\ref{conlig})  and such that the function  $\|\bm{u}_n-\tilde{\bm{u}}_n\|$ reaches a minimum, we will apply at each step in the {\tt Taylor} integrator the method of standard projection \cite{Hai} to project  $\tilde{\bm{u}}_n$  on the manifold  $\bm{F}(\bm{u})=0$. This leads to a constrained extremum problem with a Lagrangian function $L(\bm{u},\lambda):=\frac{1}{2}\|\bm{u}-\tilde{\bm{u}}\|^2-F(\bm{u}_n)^T\lambda$, where  $\lambda$  is a Lagrange multiplier. The necessary condition of  extremum and the constraint condition leads to
\begin{eqnarray}
\bm{u}_n &= \tilde{\bm{u}}_n+\nabla F(\bm{u}_n)^T\lambda\label{extrcondi1}\\
0 &= F(\bm{u}_n) \label{extrcondi2}
\end{eqnarray}
and replacing  (\ref{extrcondi1})  in  (\ref{extrcondi2}) the following nonlinear equation for $\lambda$ is obtained
\begin{equation}\label{minimiz}
    F\bigg(\tilde{\bm{u}}_n+\nabla_u F\big(\bm{u}_n(\tilde{\bm{u}}_n,\lambda)\big)^T\lambda\bigg) =0.
\end{equation}
This equation may be solved by applying the simplified Newton method. The projection stage in the algorithm spent a $3\%$ of the $0.06$ seconds-CPU employed by an Intel Core 2 Quad  processor to determine the trajectory of a photon in the time interval we are considering.

\subsection{Deformation of a wavefront by a static gravitational field}\label{subsec42}
Now we apply the method described in Section~3 to that region of a wavefront propagating along a tubular neighborhood around the $Oz^3$-axis and whose radius is $2R_\odot$. Initially, the wave surface $\mathscr{S}_0$ is flat, perpendicular to the $Oz^3$-axis and with position and velocity given in (\ref{coninic}). The wavefront $\mathscr{S}_T$ is then determined after a trip of 101 astronomical units.

\subsubsection{Curvature at a point}
\begin{figure}
\begin{center}
  \includegraphics[width=9cm,angle=0]{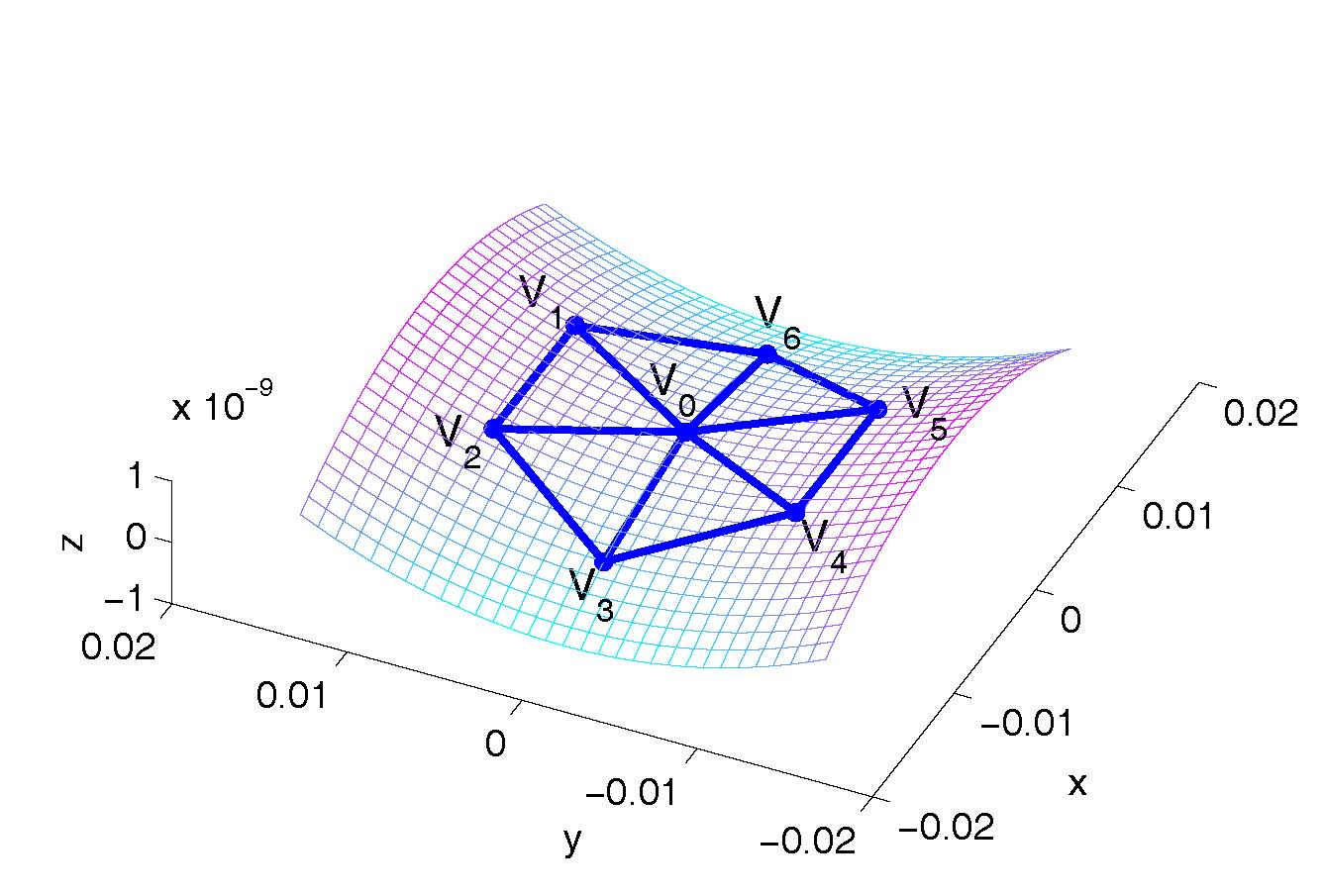}
  \caption{1-ring with vertices $[V_0,V_1,\dots,V_6]$ and corresponding least-squares quadric fitting of these vertices represented in normal coordinates centered at $V_0$.} \label{figure3}
\end{center}
\end{figure}

Firstly, we  obtain an estimation of the curvature at a point in $\mathscr{S}_T$ by using a sequence of $1$--rings with decreasing radii until a stable value of the curvature wavefront at that point is reached. Let us consider the $1$--ring of radius $r$ and centered at the point $V^{[0]}_0:=(1.0, 0.0, -21494.6550)$, determined by the three first components in  (\ref{coninic}) and whose vertices are given by
\begin{equation}\label{1anilloejemplo}
    V^{[k+1]}_0:=\big(1+r\cos(k\pi/3),\sin(k\pi/3),-21494.6550\big),\qquad (k=0,\dots 5).
\end{equation}
Now, we  apply the {\tt Taylor} algorithm for the values of tolerance and arithmetical precision  pointed out in  Subsection~\ref{subsec41}, to determine the images $V^{[k]}$ of the vertices $V^{[k]}_0$, $k=0,\dots, 6$,  by solving (\ref{edo2}) and  taking for all vertices the same initial velocity $(0.0, 0.0, 0.4307)$  (in the normalized units we are considering). After a time $T=54400$ seconds, the $1$--ring $[V^{[0]};V^{[1]},\dots, V^{[6]}]$  provides a discretized neighborhood of the point $V^{[0]}\in\mathscr{S}_T$ (see Figure~\ref{figure3}).

Applying the scheme developed in Subsection~\ref{ALFO}  we  obtain that the mean and the relative total curvatures at point $V^{[0]}$  take, for both the different values of the radius $r$ and the tolerances in the {\tt Taylor} algorithm, the values shown in Table~\ref{1anillospeq}, where one observes that for values $\text{\tt Tol} = \ncien{1.}{-20}$ and $r = R_\odot/50 $  the three first significant decimal numbers are correct.\\

\label{1anillospeq}
Table: Variation of the relative total curvature and mean curvature with respect to both the radius of the 1-ring and the tolerance used in the numerical integration of the ray equation.\\
\begin{tabular}{lllll}
\hline
&&{TOL=$\ncien{1}{-10}$} &{TOL=$\ncien{1}{-20}$}\\
\hline
Radius        & {$K_{\text{rel}}$} & {$H$}  & {$K_{\text{rel}}$} & {$H$}\\

\hline
$R_\odot/10$ & $\ncien{-0.72751}{-10}$ & $\ncien{0.15524}{-7}$ & $\ncien{-0.72752}{-10}$ & $\ncien{0.14927}{-7}$ \\
$R_\odot/20$ & $\ncien{-0.72183}{-10}$ & $\ncien{0.16073}{-7}$ & $\ncien{-.072210}{-10}$  & $\ncien{0.15350}{-7}$\\
$R_\odot/30$ & $\ncien{-0.72083}{-10}$ & $\ncien{0.16143}{-7}$ & $\ncien{-0.72110}{-10}$ & $\ncien{0.15394}{-7}$\\
$R_\odot/40$ & $\ncien{-0.72075}{-10}$ & $\ncien{0.15362}{-7}$ & $\ncien{-0.72075}{-10}$ & $\ncien{0.15407}{-7}$\\
$R_\odot/50$ & $\ncien{-0.72058}{-10}$ & $\ncien{0.15372}{-7}$ & $\ncien{-0.72059}{-10}$ & $\ncien{0.15413}{-7}$\\
$R_\odot/60$ & $\ncien{-0.72048}{-10}$ & $\ncien{0.15354}{-7}$ & $\ncien{-0.72050}{-10}$ & $\ncien{0.15415}{-7}$\\
$R_\odot/70$ & $\ncien{-0.72042}{-10}$ & $\ncien{0.15320}{-7}$ & $\ncien{-0.72044}{-10}$ & $\ncien{0.15416}{-7}$\\
$R_\odot/80$ & $\ncien{-0.72039}{-10}$ & $\ncien{0.15313}{-7}$ & $\ncien{-0.72041}{-10}$ & $\ncien{0.15416}{-7}$\\
$R_\odot/90$ & $\ncien{-0.72037}{-10}$ & $\ncien{0.15298}{-7}$ & $\ncien{-0.72039}{-10}$ & $\ncien{0.15416}{-7}$\\
$R_\odot/100$& $\ncien{-0.72036}{-10}$ & $\ncien{0.15274}{-7}$ & $\ncien{-0.72037}{-10}$ & $\ncien{0.15416}{-7}$\\
\hline
\end{tabular}

\subsubsection{Wavefront surface} \label{subsub422}
To obtain an estimation of the curvature of a region of the wavefront propagating along the $Oz^3$--axis,  we apply the  ray tracing method  with initial values on the wavefront surface described at the beginning of Subsection~\ref{subsec42}. We use a regular discretization of the wavefront surface as that described in
Subsection~\ref{subsecDiscFO}; Furthermore, taking into account the results shown in Table~\ref{1anillospeq} we choose  the length of the edges in the corresponding mesh equal to $R_\odot/50$.

In the numerical model we are studying, one assumes that the Sun is a point and we consider a hexagonal annular region on the initial wavefront similar to that shown in Figure~\ref{figure1}(b)  where the inner and outer hexagons have radii of lengths $R_\odot/25$ and  $2R_\odot$ respectively. Therefore a number of  $N:=28044$ vertices is required. The {\tt Taylor} algorithm with projection in a time interval $[0,T]$ (the time required to run a path of length equal to 101 AU) is applied to each photon located at an initial vertex; the CPU-time employed to carry out this computation is of $1544$ seconds.

For the numerical solution $\bm{u}_n$, $n=1,\dots, N$, of (\ref{edo2}), both the mean and  relative-total curvatures at each inner vertex of the mesh on the surface $\mathscr{S}_T$ are computed by applying the method described in Section~\ref{sec3}, schematized in pseudocode in the next Table.\\

\label{pseudocode}

\begin{tabular}{ll}
 {\bf Data: } $\bm{u}^*_n:=(\bm{z}^*_n,\dot{\bm{z}}^*_n), n=1,\dots N$  & \\
 {\bf for } $n=1\dots N$ {\bf do} & \\
 \qquad $\bm{u}_n :=\text{\tt Taylor}(t,\bm{u}_n^*)$                & // rays tracing   \\
 \qquad $\bm{y}_n := \text{\tt NormalCoordinates}(\bm{z}_n)$        & // see Eq. (\ref{1}) \\
 \qquad {\bf for } $i=0\dots 6$ {\bf do}                            &\\
 \qquad \qquad $\bm{y}_{n_i} := \text{\tt Ring}(\bm{y}_n)$        & // see Eq. (\ref{EntornoD}) \\
 \qquad {\bf end}                                                   & \\
 \qquad $(a_1,a_2,a_3) = {\tt LeastSquares}(\bm{y}_{n_i})$          & // see Eq. (\ref{cuadrica})\\
 \qquad $\gamma_{AB}(\bm{x}_n) := \text{\tt Metric}(a_1,a_2,a_3,\bm{x}_n)$      & // see Eq. (\ref{metrind}) \\
 \qquad $B= {\tt SecondFundamental Form}(\bm{x}_n)$                  & // see Eq. (\ref{B2})\\
 \qquad $(\lambda_1,\lambda_2) = {\tt Diagonalize}(B)$              & \\
 \qquad $(K_{\text{rel}},H) = {\tt Curvature}(\lambda_1,\lambda_2)$ & // see Eqs. (\ref{FGt},\ref{FGm})\\
 {\bf end} &\\
\end{tabular}
\\

\begin{figure}[h]
\begin{center}
 \includegraphics[width=9cm,angle=0]{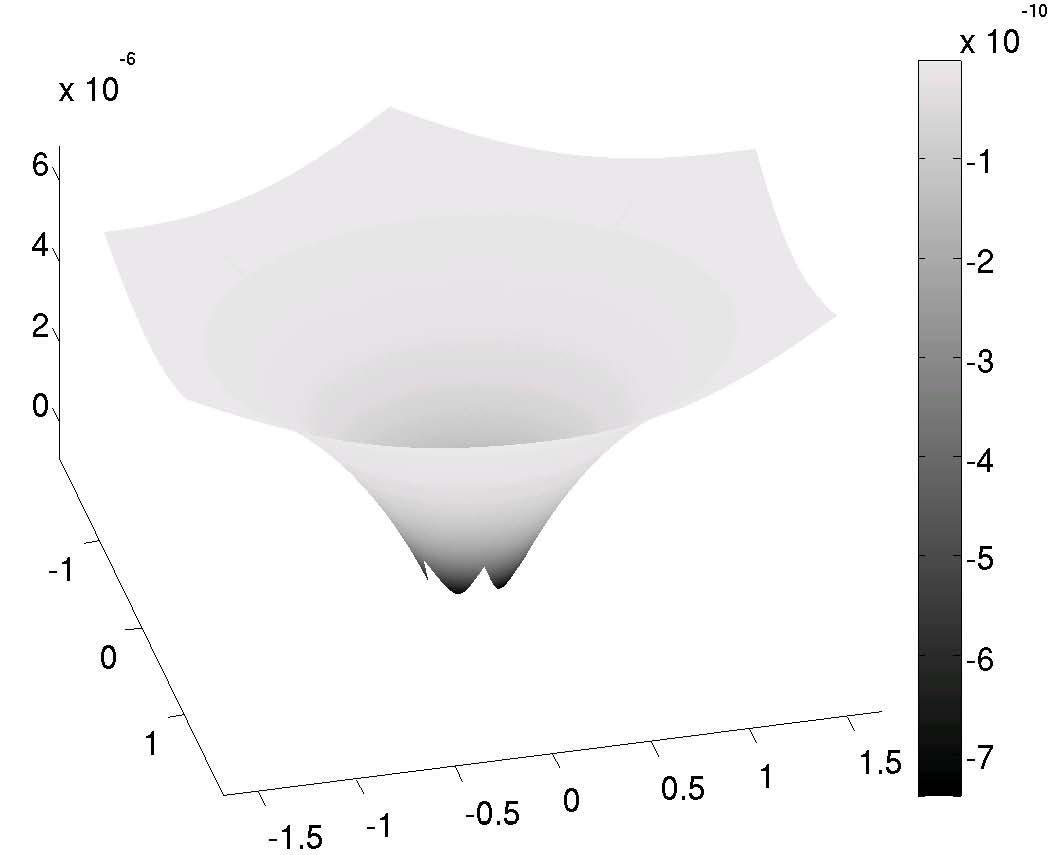}
  \caption{Wavefront surface and relative total curvature (gray scale) deformed by a spherical gravitational field (a different scale is used for the vertical axis).}\label{figure4}
\end{center}
\end{figure}

In Figure~\ref{figure4}, the surface $\mathscr{S}_T$ at the time when the wavefront arrives at the Earth is shown using a gray-scale to represent the mean curvature (note we have used a different scale on the $Oz^3$--axis). One sees in this figure that the absolute value of the mean curvature function defined on $\mathscr{S}_T$ increases as the distance between the photon and the $Oz^3$--axis decreases.

\subsection{Curvature of a wavefront in the PPN formalism}
To derive the deflection angle for light rays, instead of assuming the validity of general relativity, one may consider a more general expression for the metric generated by a spherical central body that is valid for different gravitational metric theories. In the parameterized post-Newtonian formalism, the expression of a spherically symmetric metric, written at the linearized order we are considering here, contains one parameter $\gamma$ which is usually interpreted as a modification of the curvature of the space. In the parameterized post-Newtonian formalism the total relativistic deflection angle of light rays passing near the limb of the Sun is given approximately as (see \cite{Misn}):
\begin{equation}
\Delta \phi\simeq 2(1+\gamma) 1^{\prime\prime}.75.
\end{equation}
Using very long baseline interferometry techniques, one may obtain high precision general relativistic measurements for the deflection of radio signals from distant radio sources with an accuracy at the $0.02$ percent level. In \cite{Shap} an estimation of $0.9998\pm 0.0004$ is given for the post-Newtonian parameter $\gamma$ .

\begin{figure}
\begin{center}
 \includegraphics[width=8cm,angle=0]{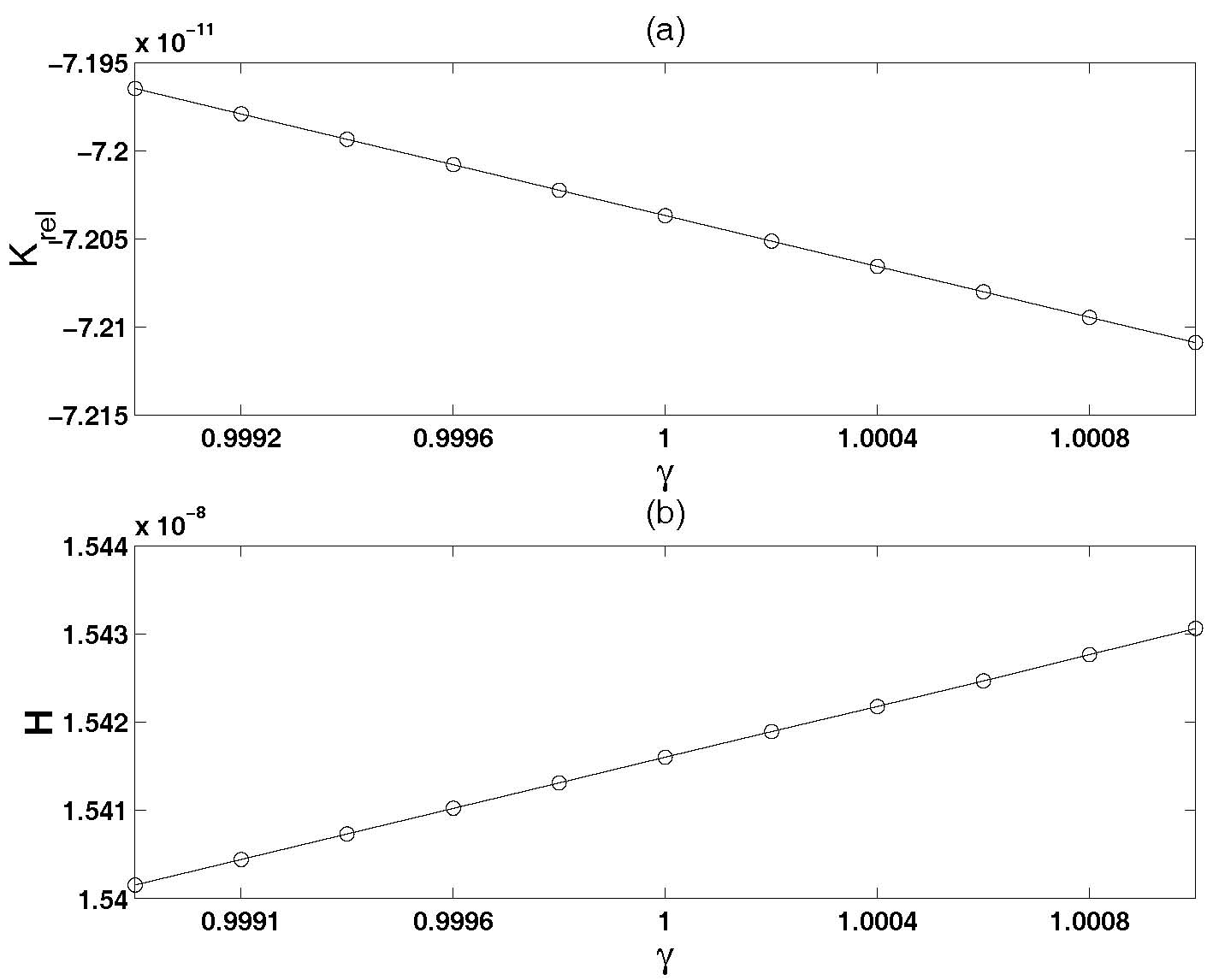}
  \caption{Relative total curvature $K_{\text{rel}}$ (a) and mean curvature $H$ (b) versus post-Newtonian parameter $\gamma$ for a wavefront numerically determined.}\label{figure5}
\end{center}
\end{figure}

Here we consider a gravitational field depending on the Eddington parameter $\gamma$ and adapt the numerical method described above to determine the dependence of the wavefront curvature at a point, corresponding to a light ray grazing the Sun with respect to this parameter. For the numerical discussion, we take a gravitational field in which only the dominant terms in the post-Newtonian metric are included, so that the metric deviations may be written as
\begin{equation}\label{PPN4}
h_{ab} =  2c^{-2}\kappa \gamma\|\bm{z}\|^{-1}\delta_{ab},\qquad
h_{a4} = 0, \qquad
h_{44} =  2c^{-2}\kappa \|\bm{z}\|^{-1}.
\end{equation}
We take values of $\gamma$ in the interval $[0.9992, 1.0012]$ in which the experimental values given in \cite{Shap} are included and consider eleven nodes in the numerical code developed above obtaining the results shown in Figure~\ref{figure5} for the curvature of the light wavefront at a point near the Earth for a light ray grazing the Sun. In this figure one observes a linear dependence of the total and mean curvature with respect to the parameter $\gamma$. For these values of $\gamma$ the variations of $K_{\text{rel}}$ and $H$ are in the second and third significant decimal digit respectively. In consequence, the numerical treatment we applied gives account of the variations of the geometric model when a change of the PPN Eddington parameter is made.

\section{Variation of the time of arrival and curvature of the wavefront}
In this section we consider the problem treated by Samuel~\cite{Sam} where the curvature of a wavefront in a gravitational field is detected by measuring the arrival times $T_a$, $a=0,1,2,3$, at four receiving stations located at points $E_a$ on an Earth  hemisphere. The arrival time differences between these stations depend on the curvature of the wavefront and are related to the volume of a parallelepiped formed by the vectors $\overrightarrow{E_0E_j}-c(T_j-T_0)\bm{n}_j$, $j=1,2,3$, where $\bm{n}:=(0,0,1)$. Samuel proposed the measurement of this non-zero volume as a new test of the general relativity.

\subsection{Volume of a tetrahedron with vertices on the wavefront surface}\label{voltetr}
\begin{figure}
\begin{center}
 \includegraphics[width=0.5\textwidth,angle=0]{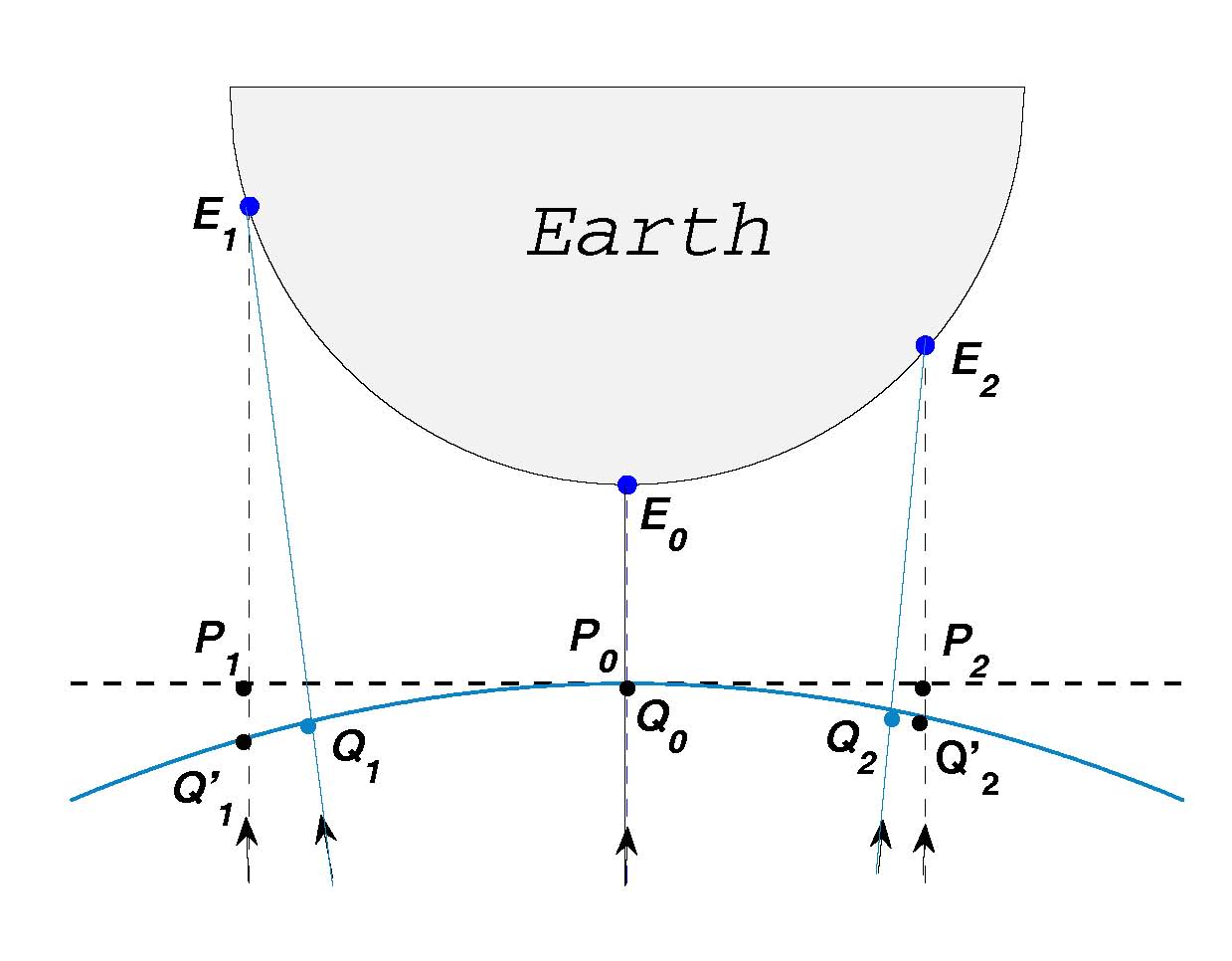}  \caption{Schematic diagram of a wavefront plane (line $P_1P_2$) and curve (line $Q_1Q_2$). Continuous lines $E_iQ_i$ represent the  light rays associated with $Q_1Q_2$ while the dashed lines $E_iP_i$ represent the associated light rays to $P_1P_2$.}\label{figure6}
\end{center}
\end{figure}

Here we compute the volume and the arrival time differences associated with a simplex determined by four points on the wavefront numerically determined above. For this, we consider a light ray reaching the reference station $E_0$ (see Figure~\ref{figure6}) at the instant of time $T_0$ and whose tangent vector in this point is $\bm{n}_0$. Let $P^*_0$ be the position of this  photon at a previous instant $T^*_0$ such that $c (T_0-T^*_0)=101\; \text{AU}$.  This position may be obtained by a backward integration of the equations of motion (\ref{edo2}) with initial values $(P_{0},-c\bm{n}_0)$. On the plane determined by $(P^*_0,\bm{n}^*_0)$, where  $\bm{n}^*_0$ is the tangent vector to the ray at the point $P^*_0$, we consider the l--ring  $[P^*_0;P^*_j]$, $j=1,\dots,6$, centered at $P^*_0$. Then, from (\ref{edo2}) and the initial values $(P^*_j, c\bm{n}^*_0)$ we determine the 1--ring $[Q_0;Q_j]$ image of $[P^*_0;P^*_j]$ by the flow (\ref{flow}) integrating again over the interval $[0,T_0-T^*_0]$.

In the numerical model we are considering, the coordinates of $P_0$ are $(2R_\odot,0,1\, \text{AU})$  and   $\bm{n}_0=(0,0,1)$. The original position $P^*_0$ and direction $\bm{n}^*_0$  of the photon are then (expressed here using only three decimal digits):
\begin{equation}
P^*_0 := (1.909, 0.0, -21494.224),\qquad \bm{n}^*_0 = (\ncien{0.183}{-5}, 0.0, 0.431).
\end{equation}
We choose as vertices $P^*_j$ the points
\begin{equation}
P^*_j= R_{(P^*_0,\boldsymbol{e}_2,\vartheta)}\big(r\cos(\pi j/3), r\sen(\pi j/3), -100 AU\big)
 \qquad j=1,\dots, 6
\end{equation}
where $R_{(P^*_0,\boldsymbol{e}_2,\vartheta)}$ represents a rotation around axis $(P^*_0,\bm{e}_2)$ through an angle  $\vartheta:=\bm{e}_3\cdot \bm{n}^*_0/\|\bm{n}^*_0\|$.

Following the same method employed in Subsection~\ref{subsub422} we perform a least-squares fitting of the data $\{Q_a\}_{a=0}^6$ to get an approximation of the wavefront surface in a neighborhood of $Q_0$ given by a quadratic surface $ \mathcal{Q} $ which, in the normal coordinates corresponding to ray $(P_0,\bm{n}_0)$, takes the form
\begin{equation}\label{cuad}
\mathcal{Q}(\bm{z}):\quad z_3=-0.1060\times 10^{-5} z_1^2+ 0.6\times 10^{-23} z_1z_2+ 0.1065\times 10^{-5} z_2^2.
\end{equation}

Now, on the Earth surface we choose four points $E_a$ with geocentric coordinates  $(0,0,-R_\oplus), (R_\oplus, 0,0), (\frac{\sqrt{3}}{2}R_\oplus, \frac{1}{2}R_\oplus,0)$ and  $(\frac{\sqrt{3}}{2}R_\oplus, -\frac{1}{2}R_\oplus,0)$ respectively, which will be transformed to normal coordinates. Assuming that the geometry of the 3--space in the vicinity of the Earth is Euclidean, one may determine the points  $Q_a\in\mathscr{Q}$ whose distances to the corresponding stations  $E_a$ are minima.

Once the  points $Q_a\in\mathscr{Q}$ corresponding to the images of the stations $E_a$ on the Earth are determined, we may compute the differences $\tau_a = \text{dist}(Q_a,Q^\prime_a)/c$ between the arrival times measured at each station under the assumptions that the wavefront surface is either a curved surface $\mathscr{Q}$  or a plane $\mathscr{P}$ determined by the pair $(P_0,\bm{n}_0)$. These differences are given, for the data provided above, by
\begin{equation}\label{desvtiempo}
\tau_0 = 0.97872481493, \quad
\tau_1 = 0.99999999032, \quad
\tau_2=\tau_3 = 0.99999999021,
\end{equation}
whereas the distances $l_{ij}:=\text{dist}(Q_i,Q_j)$, $i,j=0,1,2,3$,  between the projected  points  $Q_a$ on $\mathcal{Q}$  are (in the normalized units we are using)
\begin{eqnarray*}\label{distancias}
& l_{01} = \ncien{0.9164245}{-2}, \quad
l_{02} = \ncien{0.9164249}{-2}, \quad
l_{03} = \ncien{0.9164249}{-2}, \\
& l_{12} = \ncien{0.4743770}{-2}, \quad
l_{13} = \ncien{0.4743770}{-2}  \quad
l_{23} = \ncien{0.9164262}{-2}
\end{eqnarray*}
where $\text{dist}(A,B)$  denotes the Euclidian distance between two arbitrary points $A$ and  $B$ and $Q^\prime_a$  represents the points of minimum distance from the station $E_a$ to the plane $\mathcal{P}$. The volume of the tetrahedron determined by the points $Q_a$ is proportional to the Cayley-Menger determinant \cite{Blu} defined in terms of the lengths $l_{ij}$ of the edges and it is given by:
\begin{equation}\label{vol}
    \text{Vol} =\frac{\sqrt{2}}{24}\begin{vmatrix}
     0    &   1          &   1          &   1          &   1 \\
     1    &   0          &   l_{01}^2   &   l_{02}^2   &   l_{03}^2 \\
     1    &   l_{10}^2   &   0          &   l_{12}^2   &   l_{13}^2 \\
     1    &   l_{20}^2   &   l_{21}^2   &   0          &   l_{23}^2 \\
     1    &   l_{30}^2   &   l_{31}^2   &   l_{32}^2   &   0 \end{vmatrix},
\end{equation}
which for the lengths (\ref{distancias}) takes the value  $\text{Vol}=\ncien{0.7}{-7}$. Therefore  the  metric quadruple $(Q_a,l_{ab})$  determines a non degenerate simplex in  $\re^3$   (see for instance Saucan \cite{Sau}). This is equivalent to say that the metric quadruple $(Q_a, l_{ab}) $ is not congruent to any quadruple of points in the Euclidean plane and consequently the curvature of the wavefront surface at the point $Q_0$ is non vanishing.

\subsection{Estimation of the wavefront curvature from arrival time measurements}
Now we assume that the points $\{Q_a\}_{a=0}^3$ on the wavefront are directly  determined through measurements of arrival times. Here we consider the problem of determining an approximation of the wavefront curvature in a region far enough from the Sun (say the Earth), without resorting to the ray tracing method.

An estimation of the Gaussian curvature of the wavefront surface can be obtained using the notion of the Wald curvature  of a metric space established in the Distance Geometry (see \cite{Blu}), that in the case of 2-dimensional manifolds  agrees with the Gaussian curvature. The Wald curvature is determined  as the limit of the embedding curvatures of metric quadruples isometrically embedded in  surfaces of constant curvature (the Euclidean plane $\re^2$, the 2--sphere $\mathbb{S}^2_{\sqrt{\kappa}}$  or  the hyperbolic space $\mathbb{H}^2_{\sqrt{-\kappa}}$).

In the hyperbolic plane $\mathbb{H}^2_r$ of curvature $-1/r^2$, represented by the Blumenthal model (\cite{Blu}, p. 19) we consider the metric quadruple $(Q_j,l_{ij})$ defined in Subsection~\ref{voltetr}. The curvature of a hyperbolic plane on which there exists a quadruple congruent with $Q_j$ must fulfill both of the following conditions:
\begin{equation}\label{WaldMat}
A(r) := \begin{pmatrix}
1 & \cosh(l_{01}/r) & \cosh(l_{02}/r) & \cosh(l_{03}/r) \\
\cosh(l_{01}/r) & 1 & \cosh(l_{12}/r) & \cosh(l_{13}/r) \\
\cosh(l_{02}/r) & \cosh(l_{12}/r) & 1 & \cosh(l_{23}/r) \\
\cosh(l_{03}/r) & \cosh(l_{13}/r) & \cosh(l_{23}/r) & 1
\end{pmatrix}=0 ,
\end{equation}
and each non-zero principal minor of $A(r)$ of order $m+1$ has the sign $(- 1) ^m$ (see  \cite{Blu}, p. 274). For small values of the curvature $1/r$  the determinant $\det\big (A (r)\big)$ can be approximated by a Taylor  polynomial.  Using the symbolic processor Maple and employing numerical precision of {\tt Digits} =50 to perform the Taylor expansion we obtain that (\ref{WaldMat}) may be approximated by the following algebraic equation for $\rho:=1/r$
\begin{equation}
0.235\times 10^{-22} \rho^{10}+0.481\times 10^{-18}\rho^8-0.756\times 10^{-29}\rho^6=0
\end{equation}
having only one positive real root. This approximated solution is taken as an initial guessed solution to solve the transcendental equation (\ref{WaldMat}) by means of the command {\tt solve} in Maple. On the other hand, the conditions imposed on the sign of the principal minors are also satisfied. Therefore the Wald curvature $K_W$ associated with the chosen  quadruple $\{Q_j\}$  is
\begin{equation}
K_W := -\frac{1}{r^2}= \ncien{-0.16}{-10}.
\end{equation}
This result gives an approximation of the total curvature of the wavefront surface under the assumption that locally this surface may be identified with a hyperbolic plane in which the quadruple considered is isometrically embedded.

\subsection{Scheme of the method}
In this subsection we present an outline of the construction of the method we used above to obtain the Wald curvature corresponding to four receiving stations located at points $E_a$ and the arrival time differences $\tau_a$:
\begin{enumerate}
\item[--] Define on the wavefront  the points $P_a$ corresponding to the stations $E_a$:
\begin{equation}
\overrightarrow{OQ_a} = \overrightarrow{OE_a}+c\tau_a \bm{n},
\end{equation}
where $O$ is the coordinate origin and $\bm{n}$ is the unit vector in the direction of the light rays.
\item[--] Determine the relative distances $l_{ij}$ between the points $Q_a$:
\begin{equation}
l_{ij}^2=L_{ij}^2+2c(\tau_j-\tau_i)(\overrightarrow{OE_j}-\overrightarrow{OE_i})\cdot\bm{n}+c^2(\tau_j-\tau_i)^2,
\end{equation}
where $L_{ij}$ are the Euclidean distances between the stations $E_i$ and $E_j$.
\item[--] For the point $Q_a$ so obtained, establish the nonlinear equation (\ref{WaldMat}) in terms of the curvature of the surface.
\item[--] Solve the nonlinear equation (\ref{WaldMat}) for the unknown $r$ to obtain an estimation  of the curvature of the wavefront in a neighborhood of the station $E_0$ by means of the Wald curvature $K_W$ of this surface modeled as a hyperbolic plane.
\end{enumerate}

\section{Conclusions}
The ray tracing numerical method provides a useful tool for the description of spacelike bidimensional wavefronts  within the framework  of the general relativity. We have studied a method, based on techniques of computational geometry, that allows  to estimate the curvature properties of the surface by making a least-squares fitting of the wavefront surface by a quadric surface in the neighborhood of each point of this surface. The computation of the light rays is carried out using an algorithm based on the Taylor method for the solution of differential equations and  employing high arithmetic precision. Further, we have applied a projection at each step of the numerical integration process that allows to guarantee the fulfillment (at machine precision) of the isotropy condition  for the tangent vector to the light ray. We have also studied numerically the dependence of the curvature properties of the wavefront surface on the value of the Eddington parameter $\gamma$. On the other hand, we have employed a geometric computational approach to the study of the model proposed by Samuel as a new general relativity test, by determining  a numerical approximation of the volume corresponding to a tetrahedron formed by four points on the wavefront that reaches four receiving stations on the Earth surface. Finally, we have obtained an estimation of the Wald curvature for the wavefront in a vicinity of the Earth by using the differences of arrival time recorded at four receiving stations on the Earth.

\section{Acknowledgements}
This research was supported by the Spanish Ministerio de Educaci\'on y Ciencia, MEC-FEDER Project ESP2006-01263.



\begin{thebibliography}{00}
\bibitem{Sam} Samuel J 2004 {\it Class. Quantum Grav.\/}, {\bf 21}, L83--L88.
\bibitem{KP} Klioner S A and Peip M 2003 {\it Astron. Astrophys.\/} {\bf 410}, 1063--1074.
\bibitem{dF} de Felice F, Crosta M T, Vecchiato  A, Lattanzi M G and Bucciarelli B 2004 {\it The Astronomical Journal\/} {\bf 607}, 580--595.
\bibitem{KS} Kopeikin S and Sch\"afer G 1999 {\it Phys. Rev.\/} D {\bf 60}, 124002.
\bibitem{Lep} Le Poncin-Lafitte C, Linet B and Teyssandier P 2004 {\it Class. Quantum Grav.\/}, {\bf 21}, 4463-4483.
\bibitem{Tey} Teyssandier P and Le Poncin-Lafitte C 2008 {\it Class. Quantum Grav.\/}, {\bf 25}, 145020.
\bibitem{GS} Garimella R V  and Swartz  BK  2003 Technical Report, LA-UR-03-8240, Los Alamos National Laboratory.
\bibitem{CP} Cazals F and Pouget M 2003 {\it SGP '03: Proceedings of the 2003 Eurographics/ACM SIGGRAPH symposium on Geometry processing (Aachen, Germany)}, (Aire-la-Ville: Eurographics Association) p. 177.
\bibitem{Pet} Petitjean S 2002 {\it ACM Computing Surveys} {\bf 2}, 1--6.
\bibitem{JZ} Jorba \`A and Zou M 2005 {\it Experimental Mathematics} {\bf 14}, 99-117.
\bibitem{LL} Landau L D and Lifshitz E M 1962 {\it The Classical Theory of Fields} (Oxford: Pergamon Press).
\bibitem{Blu} Blumenthal L M 1970 {\it Theory and Applications of Distance Geometry}. (New York: Chelsea Publishing Company, 2nd edition).
\bibitem{Bru} Brumberg V A 1991 {\it Essential Relativistic Celestial Mechanics}, (Bristol: Adam Hilger).
\bibitem{BH} Bobenko A I and Hoffmann T  2003 {\it Duke Math. J.} {\bf 116} 525--566.
\bibitem{Eis} Eisenhart L P 1925 {\it Riemannian Geometry} (Princeton: Princeton University Press).
\bibitem{Br} Brewin L 1998 {\it Class. Quantum Grav}. {\bf 15} 3085--3120.
\bibitem{Sau} Saucan E 2006 Curvature --- Smooth, Piecewise-Linear and Metric, {\it What is Geometry?}, ed G Sica, ( Monza/Italy: Polimetrica International Scientific Publisher).
\bibitem{MDSB} Meyer M, Desbrum M, Schr\"oder P and Barr AH (2002) {\it Proceedings of VisMath '02} (Berlin, Germany).
\bibitem{dC} do Carmo M P 1992 {\it Riemannian Geometry} (Boston: Birkh\"auser).
\bibitem{Hai} Hairer E 2000 {\it BIT}, {\bf 40}, 726-734.
\bibitem{Misn} Misner C W, Thorne K S and Wheeler J A 1973. {\it Gravitation}, (San Francisco: Freeman).
\bibitem{Shap} Shapiro S S, Davis J L, Lebach D E and Gregory J S 2004 {\it Phys. Rev. Lett} {\bf 92}, 121101.
\end{thebibliography}
\end{document}